\documentclass[12pt,a4paper]{article}

\normalbaselines

\begin{document} 
\bibliographystyle{unsrt}
\vspace*{26mm}

  { \noindent 
 \large \bf  DISCRETE REFLECTION GROUPS AND   INDUCED  \\ REPRESENTATIONS OF  POINCARE GROUP
ON \\ THE LATTICE}\\[36pt]
\hspace*{72pt}  \noindent Miguel Lorente\\[12pt]
\hspace*{72pt}  \noindent Departamento de F\'{\i}sica \\
\hspace*{72pt}   Facultad de Ciencias \\
\hspace*{72pt}  \noindent Universidad de Oviedo \\
\hspace*{72pt}  \noindent 33007 Oviedo, Spain\\[24pt]  

\begin{abstract} 

\noindent We continue the program, presented in previous Symposia, of discretizing physical
models. In particular we calculate the integral Lorentz transformations with the help of
discrete reflection groups, and use them for the covariance of Klein-Gordon and Dirac wave
equation on the lattice. Finally we define the unitary representation of Poincar\'{e} group
on discrete momentum and configuration space, induced by integral representations of its
closed subgroup.

\end{abstract}

\vspace*{24pt} {\noindent  \bf  INTRODUCTION}\\

The assumption of a physical model on discrete space and time leads to some changes of the
mathematical equations. In particular, the symmetries of the model has to be restricted to
integral transformations acting on some vector space defined over the integers. The
space-time groups are subgroups of the set of non singular integral matrices, and the
classical and quantum laws are written with the help of difference operators. In [1] we have
developped some properties of non-compact groups acting on lattice with non-euclidean metric.
In [2] we have modify some of the postulates of quantum mechanics in order to incorporate the
hypothesis of discrete space and time. In [3] we have proposed a new scheme for the
Klein-Gordon and Dirac wave equation. The Quantum fields on the lattice is the subject of
very extensive litterature [4] and the theoretical difficulties of it have not yet been
solved.

In section 2, we review some results of discrete reflection groups that give raise to a
complete classification of n-simplex reflection groups of the spherical, euclidean or
hyperbolic type, and of the generalized hyperbolic reflection groups whose connection with
integral space-time groups is made via Coxeter groups.

In section 3 and 4 we introduce the definitions and main properties of a discrete
differential geometry in order to construct covariant expression for the Maxwell, Dirac and
Klein-Gordon difference equations.

In section 5 we apply the standard formulation of induced representations to calculate the
transformation properties of the Mackey, Wigner and covariant state functions that belong to
the carrier space for the unitary representations of Poincar\'{e} group on discrete momentum
space.

In section 6 we construct the unitary representations of Poincar\'{e} group in the discrete
coordinate space. We recover the formulas for the momentum space by the use of a discrete
Fourier transform for non-periodic functions.

\vspace*{24pt} {\noindent  \bf  REFLECTION GROUPS}\vspace*{12pt} 

Let $X$ be a metric space of dimension $n$ of the type spherical $S^n$, euclidean $E^n$ or
hyperbolic $H^n$. Let
$S$ be a side of an n-dimensional convex polyhedron $P$ in $X$. The reflection of $X$ in the
side $S$ of $P$ is the reflection of $X$ in the hyperplane spanned by
$S$. The reflections of $X$ in the sides of a finite sided n-dimensional convex polyhedron
$P$ in $X$ of finite volume generate a {\bf reflection group}. 

Let $P$ be a finite-sided n-dimensional convex polyhedron in $X$ of finite volume all of
whose dihedral angles are submultiple of $\pi$. Then the group generated by the reflections
of $X$ in the sides of $P$ is a {\bf discrete reflection group} $\Gamma$ with respect to the
polyhedron P.

In order to construct a presentation for a discrete reflection group we take all the sides
$\left\{{{S}_{i}}\right\}$ of $P$ and for each pair of indices, $i,j$, let ${k}_{ij}\equiv
{\frac{\pi }{\vartheta
\left({{S}_{i},{S}_{j}}\right)}}$ where $\vartheta \left({{S}_{i},{S}_{j}}\right)$ is the
angle between $S_i$ and
$S_j$. Let $F$ be the group generated by the sides $S_i$ and $\Gamma$ the group generated by
the reflections on
$S_i$. When $P$ has finitely many sides and finite volume the map $\psi =G\rightarrow \Gamma$
is an isomorphism where $G$ is the quotient of $F$ by the normal closure of the words
${\left({{S}_{i}{S}_{j}}\right)}^{{k}_{ij}}$. We call 
\[\left\{{{S}_{i},{\left({{S}_{i}{S}_{j}}\right)}^{{k}_{ij}}}\right\}\] a {\bf presentation}
of the discrete reflection group
$\Gamma$.

A discrete reflection group $\Gamma$ with respect to a finite-sided polyhedron $P$ of finite
volume is isomorphic to a Coxeter group $G$, that is, an abstract groups defined by a group
presentation of the form
$\left\{{{S}_{i},{\left({{S}_{i}{S}_{j}}\right)}^{{k}_{ij}}}\right\}$, where: 

\begin{enumerate}
\item[i)] the indices $i,j$ vary over some countable indexing set $J$;
\item[ii)] the exponent $k_{ij}$ is either a positive integer or $\infty$ for each $i,j$;
\item[iii)] $k_{ij}=k_{ji}$;
\item[iv)] $k_{ii}=1$, for each $i,j$;
\item[v)] $k_{ij}>1$ if $i\ne j$; and
\item[vi)] if $k_{ij}=\infty$, the term ${\left({{S}_{i}{S}_{j}}\right)}^{\infty }$ is
omitted.
\end{enumerate}

The Coxeter graph of $G$ is the labeled graph with vertices $J$ and edges

\[\left\{{\left({i,j}\right):{k}_{ij}>2}\right\}\]

Each edge $(i,j)$ is labeled by $k_{ij}$.

For simplicity, the edges with $k_{ij}=3$ are usually not labeled in a Coxeter graph.

Only some particular types of Coxeter groups have been classified. We present here some of
the most fundamental and useful Coxeter groups: the simple reflection groups and the
generalized simplex reflection groups.

Let $\Delta$ be an n-simplex in $X(=S^n, E^n {\rm or} H^n)$ all of whose diedral angles are
submultiple of $\pi$ (by n-simplex we define a convex n-dimensional polyhedron in $X$ with
exactly $n+1$ vertices). The group $\Gamma$ generated by the reflections of $X$ in the sides
of $\Delta$ is an n-simplex reflection group.

If $n=1$ then $\Delta$ is a geodesic segment in $X$, and $\Gamma$ is the dihedral group of
order $2k$ with $k>1$, where
${\frac{\pi }{k}}$ is the angle of $\Delta$. The Coxeter graph is

\setlength{\unitlength}{1mm}
\begin{center}
\begin{picture}(15,1)
\put(0,0){\circle*{1.75}}
\put(15,0){\circle*{1.75}}
\end{picture}
\hspace{2cm} or \hspace{2cm}
\begin{picture}(15,1)
\put(0,0){\circle*{1.75}}
\put(0,0){\line(1,0){15}}
\put(15,0){\circle*{1.75}}
\put(7,2){\scriptsize $k$}
\end{picture}
\end{center}

For $X=S^1$, $k$ is finite; for $X=E^1$ or $H^1$, $k=\infty$.

Assume that $n=2$. Then $\Delta$ is a triangle in $X$ whose angles ${\frac{\pi
}{a}},{\frac{\pi }{b}},{\frac{\pi }{c}}$ are submultiple of $\pi$. If we call $T(a,b,c)$ the
triangle determined by the integer numbers $(a,b,c)$, then the group
$\Gamma$ generated by the reflections in the sides of $T(a,b,c)$ is denoted by $G(a,b,c)$ and
it is called a triangle reflection group.

If $X=S^2$ the only spherical triangle reflection group have the following Coxeter graph:

\hspace*{.1cm}
$\underbrace{
\begin{picture}(30,10)
\put(0,2){\circle*{1.75}}
\put(15,2){\circle*{1.75}}
\put(30,2){\circle*{1.75}}
\end{picture}}_{G(2,2,2)}$
\hspace{2cm}
$\underbrace{
\begin{picture}(30,10)
\put(0,2){\circle*{1.75}}
\put(15,2){\circle*{1.75}}
\put(15,2){\line(1,0){15}}
\put(30,2){\circle*{1.75}}
\put(22,4){\scriptsize $c$}
\end{picture}}_{G(2,2,c)}$
\hspace{2cm}
\begin{picture}(30,10)
\put(0,2){\circle*{1.75}}
\put(0,2){\line(1,0){15}}
\put(15,2){\circle*{1.75}}
\put(15,2){\line(1,0){15}}
\put(30,2){\circle*{1.75}}
\put(15,-4.7){\makebox(0,0){\scriptsize $G$(2,3,3)}}
\end{picture}

\vspace{1cm}

\hspace*{.1cm}
$\underbrace{
\begin{picture}(30,2)
\put(0,2){\circle*{1.75}}
\put(0,2){\line(1,0){30}}
\put(15,2){\circle*{1.75}}
\put(30,2){\circle*{1.75}}
\put(22,4){\scriptsize $4$}
\end{picture}}_{G(2,3,4)}$
\hspace{2cm}
$\underbrace{
\begin{picture}(30,2)
\put(0,2){\circle*{1.75}}
\put(0,2){\line(1,0){30}}
\put(15,2){\circle*{1.75}}
\put(15,2){\line(1,0){15}}
\put(30,2){\circle*{1.75}}
\put(22,4){\scriptsize $5$}
\end{picture}}_{G(2,3,5)}$

\bigskip 

If $X=E^3$ we have the euclidean triangle reflection groups with Coxeter diagram:

\hspace*{.85cm}
$\underbrace{
\begin{picture}(15,20)
\put(0,2){\circle*{1.75}}
\put(7.5,14.5){\circle*{1.75}}
\put(15,2){\circle*{1.75}}
\put(0,2){\line(1,0){15}}
\put(0,2){\line(3,5){7.3}}
\put(15,2){\line(-3,5){7.3}}
\end{picture}}_{G(3,3,3)}$
\hspace{2,75cm}
$\underbrace{
\begin{picture}(30,2)
\put(0,2){\circle*{1.75}}
\put(15,2){\circle*{1.75}}
\put(0,2){\line(1,0){30}}
\put(30,2){\circle*{1.75}}
\put(7,4){\scriptsize $4$}
\put(22,4){\scriptsize $4$}
\end{picture}}_{G(2,4,4)}$
\hspace{2cm}
$\underbrace{
\begin{picture}(30,2)
\put(0,2){\circle*{1.75}}
\put(0,2){\line(1,0){15}}
\put(15,2){\circle*{1.75}}
\put(15,2){\line(1,0){15}}
\put(30,2){\circle*{1.75}}
\put(22,4){\scriptsize $6$}
\end{picture}}_{G(2,3,6)}$

\bigskip

If $X=H^3$ we have the hyperbolic triangle reflection groups with Coxeter diagram;

\begin{center}
$\underbrace{
\begin{picture}(30,2)
\put(0,2){\circle*{1.75}}
\put(15,2){\circle*{1.75}}
\put(0,2){\line(1,0){30}}
\put(30,2){\circle*{1.75}}
\put(7,4){\scriptsize $b$}
\put(22,4){\scriptsize $c$}
\end{picture}}_{G(2,b,c)}$
\hspace{2,75cm}
$\underbrace{
\begin{picture}(15,15)
\put(0,2){\circle*{1.75}}
\put(7.5,14.5){\circle*{1.75}}
\put(15,2){\circle*{1.75}}
\put(0,2){\line(1,0){15}}
\put(0,2){\line(3,5){7.3}}
\put(15,2){\line(-3,5){7.3}}
\put(1,8){\scriptsize $a$}
\put(7,4){\scriptsize $b$}
\put(13,8){\scriptsize $c$}
\end{picture}}_{G(a,b,c)}$
\end{center}

Let $\Gamma$ be the group generated by the reflections of $X$ in the sides of an n-simplex
$\Delta$ all of whose dihedral angles are submultiples of $\pi$. The group $\Gamma$ said to
be irreducible if and only if its Coxeter graph is connected.

The classification of all the ireducible n-simplex (spherical, euclidean and hyperbolic)
reflection groups is complete. Spherical and euclidean n-simplex reflections groups exist in
all dimensions; however, hyperbolic n-simplex reflections groups exist only for dimension
$n\le 5$.
$\left[{5}\right]$.

Another type of Coxeter groups that have been classified are the generalized simplex
reflection groups, which are defined only in $H^n$. A generalized n-simplex in $H^n$ is a
n-dimensional polyhedron with $n+1$ generalized vertices (a generalized vertex of a convex
polyhedron $P$ is either a vertex of $P$ or an ideal vertex of $P$).

The generalized hyperbolic triangle reflection groups have the following Coxeter graphs:

\setlength{\unitlength}{0.8mm}
\begin{center}
\begin{picture}(15,15)
\put(0,2){\circle*{1.75}}
\put(7.5,14.5){\circle*{1.75}}
\put(15,2){\circle*{1.75}}
\put(0,2){\line(1,0){15}}
\put(0,2){\line(3,5){7.3}}
\put(15,2){\line(-3,5){7.3}}
\put(0,8){\scriptsize $\infty$}
\put(7,-1){\scriptsize $\infty$}
\put(13,8){\scriptsize $\infty$}
\end{picture}
\hspace{2cm}
\begin{picture}(15,15)
\put(0,2){\circle*{1.75}}
\put(7.5,14.5){\circle*{1.75}}
\put(15,2){\circle*{1.75}}
\put(0,2){\line(1,0){15}}
\put(0,2){\line(3,5){7.3}}
\put(15,2){\line(-3,5){7.3}}
\put(0,8){\scriptsize $\infty$}
\put(7,-1){\scriptsize $a$}
\put(13,8){\scriptsize $\infty$}
\end{picture}
\hspace{2cm}
\begin{picture}(15,15)
\put(0,2){\circle*{1.75}}
\put(7.5,14.5){\circle*{1.75}}
\put(15,2){\circle*{1.75}}
\put(0,2){\line(1,0){15}}
\put(0,2){\line(3,5){7.3}}
\put(15,2){\line(-3,5){7.3}}
\put(0,8){\scriptsize $a$}
\put(7,-1){\scriptsize $\infty$}
\put(13,8){\scriptsize $b$}
\end{picture}
\hspace{2cm}
$a \geq b \geq 3$
\end{center}

\begin{center}
\begin{picture}(30,2)
\put(0,2){\circle*{1.75}}
\put(15,2){\circle*{1.75}}
\put(0,2){\line(1,0){30}}
\put(30,2){\circle*{1.75}}
\put(7,4){\scriptsize $\infty$}
\put(22,4){\scriptsize $\infty$}
\end{picture}
\hspace{2,75cm}
\begin{picture}(30,2)
\put(0,2){\circle*{1.75}}
\put(15,2){\circle*{1.75}}
\put(0,2){\line(1,0){30}}
\put(30,2){\circle*{1.75}}
\put(7,4){\scriptsize $b$}
\put(22,4){\scriptsize $\infty$}
\end{picture}
\hspace{2cm}
$b > 2$
\end{center}

The generalized hyperbolic n-simplex reflections groups exist only for $n\le 10$.
$\left[{6}\right]$

Some particular very interesting cases of the last hyperbolic generalized reflection groups
are the following:

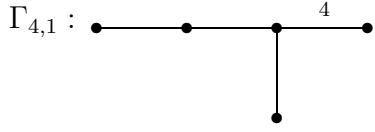
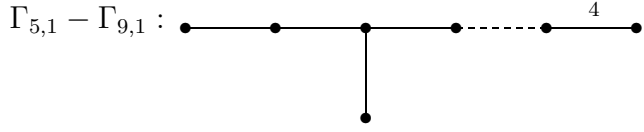
\begin{center}
$\Gamma_{2,1}:\ $
\begin{picture}(30,7)
\put(0,0){\circle*{1.75}}
\put(15,0){\circle*{1.75}}
\put(0,0){\line(1,0){30}}
\put(30,0){\circle*{1.75}}
\put(7,2){\scriptsize $4$}
\put(22,2){\scriptsize $\infty$}
\end{picture}
\hspace{1cm}
$\Gamma_{3,1}:\ $
\begin{picture}(30,7)
\put(0,0){\circle*{1.75}}
\put(15,0){\circle*{1.75}}
\put(0,0){\line(1,0){45}}
\put(30,0){\circle*{1.75}}
\put(45,0){\circle*{1.75}}
\put(22,2){\scriptsize $4$}
\put(37,2){\scriptsize $4$}
\end{picture}
\end{center}

\vspace{.2cm}

\begin{center}
$\Gamma_{4,1}:\ $
\begin{picture}(45,2)
\put(0,0){\circle*{1.75}}
\put(15,0){\circle*{1.75}}
\put(0,0){\line(1,0){45}}
\put(30,0){\circle*{1.75}}
\put(45,0){\circle*{1.75}}
\put(30,-15){\circle*{1.75}}
\put(30,0){\line(0,-1){15}}
\put(37,2){\scriptsize $4$}
\end{picture}
\hspace{.5cm}
$\Gamma_{5,1}-\Gamma_{9,1}:\ $
\begin{picture}(75,2)
\put(0,0){\circle*{1.75}}
\put(15,0){\circle*{1.75}}
\put(30,0){\circle*{1.75}}
\put(45,0){\circle*{1.75}}
\put(60,0){\circle*{1.75}}
\put(75,0){\circle*{1.75}}
\put(30,-15){\circle*{1.75}}
\put(30,0){\line(0,-1){15}}
\put(0,0){\line(1,0){45}}
\put(45,0){\dashbox{1}(15,0){}}
\put(60,0){\line(1,0){15}}
\put(67,2){\scriptsize $4$}
\end{picture}
\end{center}

\vspace{1.5cm}

If we take a cartesian basis $\left\{{{e}_{i}}\right\}\ i=0,1,\cdots,9$ a realization of the
reflections corresponding to these Coxeter groups can be given in matricial form:
\[{\Gamma }_{2,1}:\ {S}_{1}=\left({
\begin{array}{ccc}1&0&0\\ 0&0&1\\ 0&1&0\end{array} }\right),\
{S}_{2}=\left({\begin{array}{ccc}1&0&0\\ 0&1&0\\ 0&0&-1\end{array}}\right)\]
\[{S}_{3}=\left({\begin{array}{ccc}3&2&2\\ -2&-1&-2\\ -2&-2&-1\end{array}}\right)\]
\[{\Gamma }_{3,1}:\ {S}_{1}=\left({\begin{array}{cccc}1&0&0&0\\ 0&0&1&0\\ 0&1&0&0\\
0&0&0&1\end{array}}\right),\ {S}_{2}=\left({\begin{array}{cccc}1&0&0&0\\ 0&1&0&0\\ 0&0&0&1\\
0&0&1&0\end{array}}\right),\ {S}_{3}=\left({\begin{array}{cccc}1&0&0&0\\ 0&1&0&0\\ 0&0&1&0\\
0&0&0&-1\end{array}}\right)\]
\[{S}_{4}=\left({\begin{array}{cccc}2&1&1&1\\ -1&0&-1&-1\\ -1&-1&0&-1\\
-1&-1&-1&0\end{array}}\right)\]
\[{\Gamma }_{n,1}:\ {S}_{i}={x}_{i}\leftrightarrow {x}_{i+1}\ \ i=1,2,\ \cdots,\ n-1\ ;\
n=4,5,\cdots,9.\]
\[{S}_{n}={x}_{n}\leftrightarrow {-x}_{n}\]
\[{S}_{n+1}\ =\ \left({\begin{array}{cc}\begin{array}{cccc}2&1&1&1\\ -1&0&-1&-1\\
-1&-1&0&-1\\ -1&-1&-1&0\end{array}&0\\ 0&{I}_{n-3}\end{array}}\right)\]

It can be checked that these reflections satisfy the presentation of the Coxeter groups
\[\left\{{{S}_{i},{\left({{S}_{i}{S}_{j}}\right)}^{{k}_{ij}}}\right\}\]
corresponding to the Coxeter graphs ${\Gamma }_{n,1},\left({n=2,3,\cdots,9}\right)$

The groups ${\Gamma }_{n,1}$ have a very important conection with the Lorentian lattices
${\Lambda }_{n}$ or the set of points with integral components with respect to a Cartesian
basis $\left\{{{e}_{0},{e}_{1},\cdots,{e}_{n}}\right\}$ equipped with the bilinear form
\[-{x}_{0}^{2}+{x}_{1}^{2}+\cdots+{x}_{n}^{2}\]

The set of all proper Lorentz transformations that leave the Lorentian lattice ${\Lambda
}_{n}$ invariant is generated by the elements of the Coxeter group ${\Gamma
}_{n,1},\left({n=2,\cdots,9}\right)$, the proof is given by Kac
$\left[{7}\right]$. Similarly Coxeter has proved $\left[{8}\right]$ that all integral Lorentz
transformations (including reflections) are obtained combining the operation of permuting the
spacial coordinates $x_1, x_2, x_3$ and changing the signs of the coordinates $x_0, x_1, x_2,
x_3,$ together with the transformation
\[\left({\begin{array}{cccc}2&-1&-1&-1\\ -1&0&-1&-1\\ -1&-1&0&-1\\
-1&-1&-1&0\end{array}}\right)\]

We have arrived at the same result in the case of ${\Gamma }_{2,1}$ using the isomorphism
between ${\rm SO}\left({2,1}\right)\cap {\rm GL}\left({3,{\cal Z}}\right)$ and ${\rm
GL}\left({2,{\cal Z}}\right)$ [9]

\vspace*{24pt} {\noindent  \bf  A DIFFERENCE CALCULUS OF SEVERAL INDEPENDENT 
VARIABLES}\vspace*{12pt} 

\setcounter{section}{3}\setcounter{equation}{0}
Given a function of one independent variable the forward and backward differences are defined
as
\[ \Delta f(x) \equiv f(x + \Delta x) - f(x) \quad , \quad \nabla f(x) \equiv f(x) - f(x -
\Delta x)\]

Similarly, we can define the forward and backward promediate operator
\[\widetilde{\Delta }f(x) \equiv {1 \over 2}\left\{{f(x+\Delta x) + f(x)}\right\} \quad ,
\quad
\widetilde{\nabla }f(x) \equiv {1 \over 2}\left\{{f(x-\Delta x) + f(x)}\right\}\]

Hence the difference or promediate of the product of two functions follows:
\begin{eqnarray}
\Delta \left\{{f(x) g(x)}\right\} &=& \Delta f(x) \widetilde{\Delta } g(x) +
\widetilde{\Delta } f(x) \Delta g(x) \\
\widetilde{\Delta } \left\{{f(x) g(x)}\right\} &=& \widetilde{\Delta } f(x)
\widetilde{\Delta } g(x) + {1 \over 4}\Delta f(x) \Delta g(x)
\end{eqnarray}

This calculus can be enlarged to functions of several independent variables. We use the
following definitions:
\begin{eqnarray*} {\Delta }_{x} f(x,y) &\equiv & f(x+\Delta x,y) - f(x,y) \\ {\Delta }_{y}
f(x,y) &\equiv & f(x,y+\Delta y) - f(x,y) \\ {\widetilde{\Delta }}_{x} f(x,y) &\equiv &
\frac{1}{2} \left\{{f (x+\Delta x,y) + f(x,y)}\right\} \\ {\widetilde{\Delta }}_{y} f(x,y)
&\equiv &{1 \over 2} \left\{{f (x,y+\Delta y) + f(x,y)}\right\}
\\
\Delta f(x,y) &\equiv & f (x+\Delta x , y+\Delta y) - f(x,y) \\
\widetilde{\Delta } f(x,y) &\equiv & {1 \over 2} \left\{{f (x+\Delta x , y+\Delta y) +
f(x,y)}\right\}
\end{eqnarray*}

These definitons can be easily generalized to more independent variables but for the sake of
brevity we restrict ourselves to two independent variables. From the last definitions it can
be proved the following identities: 
\begin{eqnarray}
\Delta f(x,y) &=& {\Delta }_{x} {\widetilde{\Delta }}_{y} f(x,y) + {\widetilde{\Delta }}_{x}
{\Delta }_{y} f(x,y) \\
\widetilde{\Delta } f(x,y) &=& {\widetilde{\Delta }}_{x} {\widetilde{\Delta }}_{y} f(x,y) +
{1 \over 4}{\Delta }_{x} {\Delta }_{y}f(x,y)
\end{eqnarray}

\noindent\hspace*{5mm}Given a vectorial space $V^n$ over $\cal Z$ we can define a real-valued
linear function over $\cal Z$
\begin{equation} f\left({u}\right) \equiv \left\langle{\omega , u}\right\rangle \quad u\in {
V}^{ n}
\end{equation}

The forms $\omega$ constitue a vectorial linear space (dual space) ${}^*{V}^{n}$, and can be
expanded in terms of a basis ${\omega }^{\alpha }$
\vspace{-2mm}$$\omega = {\sigma }_{\alpha } {\omega }^{\alpha }$$

The basis $e_{\beta}$ of $V^n$ and $\omega^{\alpha}$ of ${}^*{V}^{n}$ can be contracted in
the following way
\begin{equation}
\left\langle{{\omega }^{\alpha } , {e}_{\beta }}\right\rangle {\delta }_{\beta}^{\alpha } 
\end{equation}
hence
\begin{equation}
\left\langle{\omega , {e}_{a}}\right\rangle = {\sigma }_{\alpha } \quad , \quad 
\left\langle{{\omega }^{\beta } , u}\right\rangle = {u}^{\beta } \quad , \quad
\left\langle{\omega , u}\right\rangle = {\sigma }_{\alpha } {u}^{\alpha }
\end{equation}
 with $u={u}^{\alpha }{e}_{\alpha }$.

If we take ${\omega }^{\beta } = \Delta {x}^{\beta }$ as coordinate basis for the linear
forms we can construct discrete differential forms (a discrete version of the continuous
differential forms) [10]

A particular example of this discrete form is the total difference operator (3,3) of a
function of several discrete variables written in the following way:
\begin{equation}
\Delta f(x,y) = \left({{{\Delta }_{x} {\widetilde{\Delta }}_{y} f \over \Delta x}}\right)
\Delta x +
\left({{{\widetilde{\Delta }}_{x} {\Delta }_{y} f \over \Delta y}}\right) \Delta y
\end{equation}
For these discrete forms we can define the exterior product of two form $\sigma$ and
$\rho$
$$\rho \wedge \sigma = -\sigma \wedge \rho $$ 
which is linear in both arguments.

For the coordinate basis we also have
$$\Delta x \wedge \Delta y = -\Delta y \wedge \Delta x$$
With the help of this exterior product we can construct a second order discrete differential
form or 2-form, namely 
\begin{equation}
\rho \wedge \sigma = {\rho }_{\alpha } {\Delta x}^{\alpha } \wedge {\sigma }_{\alpha }
{\Delta x}^{\beta } = {1
\over 2} \left({{\rho }_{\alpha } {\sigma }_{\beta } - {\rho }_{\beta }{\sigma }_{\alpha
}}\right) {\Delta x}^{\alpha }
\wedge {\Delta x}^{\beta }
\equiv {\sigma }_{\alpha
\rho } {\Delta x}^{\alpha } \wedge {\Delta x}^{\beta }
\end{equation} 
where ${\sigma }_{\alpha \rho }$ is an antisymmetric tensor. Similarly we can define a
discrete p-form in a $n$-dimensional space $(p<n)$
$$\sigma = {1 \over p!} {\sigma }_{i_1}{}_{i_2}\ldots {}_{i_p} {\Delta x}^{{i}_{1}}
\wedge {\Delta x}^{{i}_{2}} \ldots \wedge {\Delta x}^{{i}_{p}}$$ 
where ${\sigma }_{i_1}{}_{i_2}\ldots {}_{i_p}$ is a totally antisymmetric tensor

The dual of a $p$-form in a $n$-dimensional space is the $(n-p)$-form $^*\alpha$ with
components
$$ {\left({{}^*\alpha }\right)}_{{k}_{1} {k}_{2}\ldots {k}_{n-p}} = {1 \over p!} {\alpha
}^{{i}_{1} {i}_{2}\ldots {i}_{p}} {\left.{\varepsilon }\right.}_{{i}_{1} {i}_{2}\ldots
{i}_{p}{k}_{1}\ldots {k}_{n-p}}$$ 
where $\varepsilon$ is the $n$-dimensional Levy-Civitt\'a totally antisymmetric
tensor $\left({{\varepsilon }_{1 2 3 \ldots } \equiv 1}\right)$


\vspace*{24pt} {\noindent  \bf  EXTERIOR CALCULUS AND LORENTZ TRANSFORMATIONS}\vspace*{12pt} 

\setcounter{section}{4}\setcounter{equation}{0}
Given a 1-form in a two-dimensional space
$$\omega = a(x,y) \Delta x + b(x,y) \Delta y$$ we can define the exterior difference, in the
similar way as the exterior derivative, namely, 
\begin{eqnarray}
\Delta \omega &\equiv& \Delta a \wedge \Delta x + \Delta b \wedge \Delta y \nonumber \\ &=&
\left({{{\Delta }_{x}{\widetilde{\Delta }}_{y} a \over \Delta x} \Delta x +
{{\widetilde{\Delta }}_{x} {\Delta }_{y} a
\over \Delta y} \Delta y}\right) \wedge \Delta x + \left({{{\Delta }_{x} {\widetilde{\Delta
}}_{y} b
\over \Delta x} \Delta x + {{\widetilde{\Delta }}_{x} {\Delta }_{y} b \over \Delta y}
\Delta y}\right) \wedge \Delta y \nonumber \\ &=& \left({{{\Delta }_{x} {\widetilde{\Delta
}}_{y} b \over \Delta x} - {{\widetilde{\Delta }}_{x} {\Delta }_{y} a
\over \Delta y}}\right) \Delta x \wedge \Delta y
\end{eqnarray} where in the last expression we have used the properties of the exterior
product.

This definition of exterior difference can be easily written for 1-form in
$n$-dimensional space.

Given a 2-form in a 3-dimensional space,
\begin{equation}
\omega = a (x,y,z) \Delta y \wedge \Delta z + b(x,y,z) \Delta z \wedge \Delta x + c (x,y,z)
\Delta x \wedge \Delta y
\end{equation} we can also define the exterior difference as:
\begin{eqnarray}
\Delta \omega &=& \Delta a \wedge \Delta y \wedge \Delta z + \Delta b \wedge \Delta z
\wedge \Delta x + \Delta c \wedge \Delta x \wedge \Delta y \nonumber \\ &=&
\left({{{\Delta }_{x} {\widetilde{\Delta }}_{y} {\widetilde{\Delta }}_{z} a \over \Delta x} +
{{\widetilde{\Delta }}_{x} {\Delta }_{y} {\widetilde{\Delta }}_{z} b \over \Delta y} +
{{\widetilde{\Delta }}_{x} {\widetilde{\Delta }}_{y} {\Delta }_{z} c \over \Delta z}}\right)
\Delta x \wedge \Delta y \wedge \Delta z
\end{eqnarray}

The exterior derivative applied to the product of a 0-form (scalar function f) and a 1-form
$(\omega = a \Delta x + b
\Delta y)$ is
\begin{equation}
\Delta \left({f\omega }\right) = \widetilde{\Delta }f \Delta \omega + \Delta f \wedge
\widetilde{\Delta }\omega 
\end{equation} where $\widetilde{\Delta }f$ is expressed in (3.4) and $\widetilde{\Delta
}\omega =
\widetilde{\Delta }a \Delta x + \widetilde{\Delta }b \Delta y$

The exterior difference of the product of two 1-forms is easily obtained
\begin{equation}
\Delta \left\{{{\omega }_{1} \wedge {\omega }_{2}}\right\} = \Delta {\omega }_{1}
\wedge
\widetilde{\Delta } {\omega }_{2} - \widetilde{\Delta } {\omega }_{1} \wedge \Delta {\omega
}_{2}
\end{equation}

The exterior difference of the product of a p-form $\rho$ and a q-form $\sigma$ is
\begin{equation}
\Delta \left\{{\rho \wedge \sigma }\right\} = \Delta \rho \wedge \widetilde{\Delta }
\sigma + {\left({-1}\right)}^{p} \widetilde{\Delta } \rho \wedge \Delta \sigma
\end{equation}

Finally for any p-form $\omega$ we have
\begin{equation} {\Delta }^{2} \omega = \Delta \left({\Delta \omega }\right) = 0
\end{equation}

Some examples:

From the Faraday 2-form
$\bf \mit {\bf F}\rm ={\frac{1}{2}}{F}_{\mu \nu }\Delta {x}^{\mu }\wedge \Delta {x}^{\nu }$
we write down one set of Maxwell difference equations
$$
\Delta {\bf F} = \Delta \left({\Delta {\bf A}}\right) = 0
$$
\begin{eqnarray}
\left({{{\Delta }_{x} {\widetilde{\Delta }}_{y} {\widetilde{\Delta }}_{z} {\widetilde{\Delta
}}_{t} {B}_{x} \over
\Delta x} + {{\widetilde{\Delta }}_{x} {\Delta }_{y} {\widetilde{\Delta }}_{z}
{\widetilde{\Delta }}_{t} {B}_{y}
\over \Delta y} + {{\widetilde{\Delta }}_{x} {\widetilde{\Delta }}_{y} {\Delta }_{z}
{\widetilde{\Delta }}_{t} {B}_{z}
\over \Delta z}}\right) \Delta x \wedge \Delta y \wedge \Delta z \nonumber
\\ + \left({{{\widetilde{\Delta }}_{x} {\widetilde{\Delta }}_{y} {\widetilde{\Delta }}_{z}
{\Delta }_{t} {B}_{x}
\over \Delta t} + {{\widetilde{\Delta }}_{x} {\Delta }_{y} {\widetilde{\Delta }}_{z}
{\widetilde{\Delta }}_{t} {E}_{z}
\over \Delta y} - {{\widetilde{\Delta }}_{x} {\widetilde{\Delta }}_{y} {\Delta }_{z}
{\widetilde{\Delta }}_{t} {E}_{y}
\over \Delta z}}\right) \Delta t \wedge \Delta y \wedge
\Delta z  \nonumber \\ +\left({{{\widetilde{\Delta }}_{x} {\widetilde{\Delta }}_{y}
{\widetilde{\Delta }}_{z} {\Delta }_{t} {B}_{y} \over \Delta t} + {{\widetilde{\Delta }}_{x}
{\widetilde{\Delta }}_{y} {\Delta }_{z} {\widetilde{\Delta }}_{t} {E}_{x}
\over \Delta z} - {{\Delta }_{x} {\widetilde{\Delta }}_{y} {\widetilde{\Delta }}_{z}
{\widetilde{\Delta }}_{t} {E}_{z}
\over \Delta x}}\right) \Delta t \wedge \Delta z \wedge
\Delta x  \nonumber \\ +\left({{{\widetilde{\Delta }}_{x} {\widetilde{\Delta }}_{y}
{\widetilde{\Delta }}_{z} {\Delta }_{t} {B}_{z} \over \Delta t} + {{\Delta }_{x}
{\widetilde{\Delta }}_{y} {\widetilde{\Delta }}_{z} {\widetilde{\Delta }}_{t} {E}_{y}
\over \Delta x} - {{\widetilde{\Delta }}_{x} {\Delta }_{y} {\widetilde{\Delta }}_{z}
{\widetilde{\Delta }}_{t} {E}_{x}
\over \Delta y}}\right) \Delta t \wedge
\Delta x \wedge \Delta y
\end{eqnarray} from the Maxwell 2-dual form ${}^*{\bf F}$ we get the other set of Maxwell
equations:
$$
\Delta {}^*{\bf F} = 4\pi {}^*{\bf J}
$$

where $\bf \mit {\bf J}\rm ={J}_{\mu }\Delta {x}^{\mu }$ is the charge-current 1-form.

Taking the exterior derivative of the last equation we get an other example of ${\Delta }^{2}
=0$.
\begin{eqnarray}
\left({{{\widetilde{\Delta }}_{x} {\widetilde{\Delta }}_{y} {\widetilde{\Delta }}_{z} {\Delta
}_{t}
\rho \over \Delta t} + {{\Delta }_{x} {\widetilde{\Delta }}_{y} {\widetilde{\Delta }}_{z}
{\widetilde{\Delta }}_{t} {J}_{x}
\over \Delta x} + {{\widetilde{\Delta }}_{x} {\Delta }_{y} {\widetilde{\Delta }}_{z}
{\widetilde{\Delta }}_{t} {J}_{y}
\over \Delta y} + {{\widetilde{\Delta }}_{x} {\widetilde{\Delta }}_{y} {\Delta }_{z} {\Delta
}_{t} {J}_{z}
\over \Delta z}}\right) \cdot  \nonumber \\
\cdot  \Delta t \wedge \Delta x \wedge \Delta y \wedge \Delta z = 0
\end{eqnarray}

Note that the coefficient of the difference form is the discrete version of the continuity
equation.

From a scalar function we get the wave equations in terms of difference operators, namely, 
\begin{equation} -{}^*\Delta {}^*\Delta \phi \equiv
\raisebox{0,5ex}{\fbox{\rule{0mm}{0,5ex}\hspace*{0,5ex}}}\; \phi 
\end{equation} where \raisebox{0,5ex}{\fbox{\rule{0mm}{0,5ex}\hspace*{0,5ex}}} is the
discrete d'Alambertian operator:
\begin{eqnarray}
\left\{ - {\widetilde{\nabla }}_{x}{\widetilde{\nabla }}_{y}{\widetilde{\nabla }}_{z}{\nabla
}_{t}\left({{\widetilde{\Delta }}_{x}{\widetilde{\Delta }}_{y}{\widetilde{\Delta
}}_{z}{\Delta }_{t}}\right)+{\nabla }_{x}{\widetilde{\nabla }}_{y}{\widetilde{\nabla
}}_{z}{\widetilde{\nabla }}_{t}\left({{\Delta }_{x}{\widetilde{\Delta
}}_{y}{\widetilde{\Delta }}_{z}{\widetilde{\Delta }}_{t}}\right)
\right. \nonumber \\
\left. +{\widetilde{\nabla }}_{x}{\nabla }_{y}{\widetilde{\nabla }}_{z}{\widetilde{\nabla
}}_{t}\left({{\widetilde{\Delta }}_{x}{\Delta }_{y}{\widetilde{\Delta
}}_{z}{\widetilde{\Delta }}_{t}}\right)+{\widetilde{\nabla }}_{x}{\widetilde{\nabla
}}_{y}{\nabla }_{z}{\widetilde{\nabla }}_{t}\left({{\widetilde{\Delta
}}_{x}{\widetilde{\Delta }}_{y}{\Delta }_{z}{\widetilde{\Delta }}_{t}}\right)\right\}\phi
\left({ xyzt}\right) =0
\end{eqnarray}

In order to compute the Lorentz transformation of the discrete differential forms we start
with the coordinate-independent nature of 1-form

\begin{equation}
\omega ={\omega }_{\mu }\Delta { x}^{\mu }
\end{equation} where the $\Delta x^{\mu }$ are the space-time intervals in Minskowski
space-time. From

\begin{equation}
\Delta { x}^{\mu \prime} ={\Lambda }_{\nu }^{\mu \prime}\Delta { x}^{\nu }
\end{equation} where ${\Lambda }_{\nu }^{\mu \prime}$ is a global Lorentz transformation, and
from the coordinate-free expresion for $\omega$ we get

\begin{equation} {\omega }_{\mu \prime}={\omega }_{\nu }{\Lambda }_{\mu \prime}^{\nu }
\end{equation} Recall that ${\Lambda }_{\mu \prime}^{\nu }{\Lambda }_{\rho }^{\mu
\prime}={\delta }_{\rho }^{\nu }$

From the total difference of a function of several variables $f\left({x,y,z,t}\right)$

\begin{equation}
\Delta f={{\Delta }_{x}{\widetilde{\Delta }}_{y}{\widetilde{\Delta }}_{z}{\widetilde{\Delta
}}_{t}f \over \Delta x}\Delta x+{{\widetilde{\Delta }}_{x}{\Delta }_{y}{\widetilde{\Delta
}}_{z}{\widetilde{\Delta }}_{t}f \over \Delta y}\Delta y+{{\widetilde{\Delta
}}_{x}{\widetilde{\Delta }}_{y}{\Delta }_{z}{\widetilde{\Delta }}_{t}f \over \Delta z}\Delta
z+{{\widetilde{\Delta }}_{x}{\widetilde{\Delta }}_{y}{\widetilde{\Delta }}_{z}{\Delta }_{t}f
\over
\Delta t}\Delta t
\end{equation} it follows that the coefficients of the 1-forms, namely,

\begin{equation}
\left({{{\Delta }_{x}{\widetilde{\Delta }}_{y}{\widetilde{\Delta }}_{z}{\widetilde{\Delta
}}_{t}f
\over \Delta x},{{\widetilde{\Delta }}_{x}{\Delta }_{y}{\widetilde{\Delta
}}_{z}{\widetilde{\Delta }}_{t}f \over
\Delta y},{{\widetilde{\Delta }}_{x}{\widetilde{\Delta }}_{y}{\Delta }_{z}{\widetilde{\Delta
}}_{t} \over \Delta z},{{\widetilde{\Delta }}_{x}{\widetilde{\Delta }}_{y}{\widetilde{\Delta
}}_{z}{\Delta }_{t} \over \Delta t}}\right)
\end{equation}  transform covariantly like the coefficients ${\omega }_{\mu }$ of (4.1).

The same technic can be applied to components of discrete p-forms. 


\vspace*{24pt}{\noindent  \bf  UNITARY REPRESENTATIONS OF THE DISCRETE POINCARE
\\ GROUP IN MOMENTUM SPACE}\vspace*{12pt} 

\setcounter{section}{5}\setcounter{equation}{0}
Let $\cal P_{+}$ be the integral proper inhomogeneous Lorentz group or discrete Poincar\'{e}
group, which is homomorphic to the semidirect product of the integral subgroups of
$SL\left({2,{\cal C}}\right)$ and ${T}_{4}$ the transformation group on the Minkowski
lattice. The multiplication law for the Poincar\'{e} group is

\begin{equation} \left\{{{a}_{2},{\Lambda }_{2}}\right\}\left\{{{a}_{1},{\Lambda
}_{1}}\right\}=\left\{{{a}_{2}+{\Lambda }_{2}{a}_{1},{\Lambda }_{2}{\Lambda
}_{1}}\right\}\end{equation} where ${\Lambda }_{1},{\Lambda }_{2}$ are two integral Lorentz
transformations described in section 2 and $a_1, a_2$ are two set of four integer numbers
that define the parameters of the translation groups $T_4$ in Minkowski space.

In order to construct the unitary representation of the Poincar\'{e} group we use the
standard method of induced representation [11]. All the properties of these representations
can be translated into the language of integral representations. It is well known that the
irreducible representations of the inhomogeneous space-time groups can be realized on
different kind of states such as Mackey, Wigner or covariant states.

Consider now the representations acting on the Wigner states, corresponding to massive
particles. Following standard procedure we need:

i) the unitary representation of the translation group

\begin{equation} U\left({a}\right)=\prod\limits_{\mu \ =\ 0}^{3}
{\left({{\frac{1+{\frac{1}{2}}i\varepsilon {P}_{\mu }}{1-{\frac{1}{2}}i\varepsilon {P}_{\mu
}}}}\right)}^{{a}_{\mu }}\end{equation} $k_{\mu }$ being the discrete momentum,
$a=(a_0,a_1,a_2,a_3)$ the parameters of the group.

ii) The unitary representation of the Wigner little group (in our case $SU(2)$) corresponding
to integral rotations in 3-dimensional space. They are only 24 different elements of this
type.

iii) The representative momentum is $\left({m,0,0,0}\right)\equiv {\stackrel{\circ}{P}}_{\mu
}$ and the orbit generated by this vector is given by the boost

\begin{equation} {P}_{\mu }={\Lambda }_{\mu }^{\nu }{\stackrel{\circ}{P}}_{\nu }\end{equation}
$P_{\nu }$ is defined on the Minkowski lattice:

\begin{equation} {P}_{\mu }=m\left({{\frac{\Delta t}{\sqrt {\Delta {t}^{2}-{\left({\Delta
\vec{x}}\right)}^{2}}}},\ {\frac{\Delta\vec{x}}{\sqrt {\Delta {t}^{2}-{\left({\Delta
\vec{x}}\right)}^{2}}}}}\right)\end{equation} the points
$(t,\vec{x})$ lying in the integral hyperboloid generated by the reflection $S_4$ of ${\Gamma
}_{3,1}$ (see section 2).

With these ingredients we can write down the transformation properties of the Wigner
functions in momentum space under the transformations of the restricted Poin\-car\'{e} group,
namely [Ref. 11, formula 16.2]

\begin{equation} U\left({a,\Lambda }\right)\psi \left({P}\right)=\prod\limits_{\mu =0}^{3}
{\left({{\frac{1+{\frac{i}{2}}\varepsilon {P}_{\mu }}{1-{\frac{i}{2}}\varepsilon {P}_{\mu
}}}}\right)}^{{a}_{\mu }}D\left({SU\left({2}\right)}\right)\psi\left({{\Lambda
}^{-1}P}\right)\end{equation} 

The integral representation of the $SU(2)$ group can be given by the use of Cayley
parametrization

\[A\ =\ {\frac{1}{\det}}\left( {\begin{array}{cc} {n}_{0}+i{n}_{3}&{n}_{2}+i{n}_{1}\\
{-n}_{2}+i{n}_{1}&{n}_{0}-i{n}_{3}
\end{array}}
\right)={\frac{1}{\det}}\left({{n}_{0}1+i{n}_{1}{\sigma }_{1}+i{n}_{2}{\sigma
}_{2}+i{n}_{3}{\sigma }_{3}}\right)\]  with
$\det={n}_{0}^{2}+{n}_{1}^{2}+{n}_{2}^{2}+{n}_{3}^{2}=1\ \mbox{or}\ 2$

For representation of higher dimension we have to substitute the generators
$\vec{\sigma}$ by the corresponding representations of $\vec{\sigma}$.

For the covariant states the unitary representations of the Poincar\'{e} group restricted to
integral transformations is given by the following transformations [see Ref. 11, formula 17.1]

\begin{equation} U\left({a,\Lambda }\right)\psi \left({P}\right)=\prod\limits_{\mu =0}^{3}
{\left({{\frac{1+{\frac{1}{2}}i\varepsilon {P}_{\mu }}{1-{\frac{1}{2}}i\varepsilon {P}_{\mu
}}}}\right)}^{{a}_{\mu }}D\left({\Lambda }\right)\psi \left({{\Lambda
}^{-1}P}\right)\end{equation} where $D\left({\Lambda }\right)$ is an integral representation
of the restricted Lorentz group.

In our case it is a representation of the subgroup $SL\left({2,C}\right)\cap
GL\left({2,Z}\right)$ which is homomorphic to the integral Lorentz group as described in
section 2.

In order to construct covariant functions that transform under irreducible representation of
the Poincar\'{e} group we need a subsidiary condition on the functions:

\begin{equation} Q\psi \left({\stackrel{\circ}{P}}\right)=\psi
\left({\stackrel{\circ}{P}}\right)\end{equation} where $Q$ is a projection operator that
restricts the components of the vector $\psi \left({\stackrel{\circ}{P}}\right)$ to one
representation of unique spin.

For example for the Dirac representation of $SL\left({2,C}\right)$ which contains
${D}^{1/2}\left({SU\left({2}\right)}\right)$ twice, we take

\begin{equation} Q={\frac{1}{2}}\left({1\pm \beta }\right),\qquad \beta
=\left({\begin{array}{cc}I&0\\ 0&-I\end{array}}\right)\end{equation} 

Taking $\stackrel{\circ}{P}=\left({{m}_{0},0,0,0}\right)$ this equation leads to the wave
equation in the rest system

\begin{equation} \beta {\psi }^{\left({\pm }\right)}\left({\stackrel{\circ}{P}}\right)=\pm
{\psi }^{\left({\pm }\right)}\left({\stackrel{\circ}{P}}\right)\end{equation} 

If we transform this equation to an arbitrary inertial system we calculate

\[Q\left({P}\right)=F\left({P}\right)Q{F}^{-1}\left({P}\right)\]

\[\psi \left({P}\right)=F\left({P}\right)\psi \left({\stackrel{\circ}{P}}\right)\]  with 
\[F\left({P}\right)={\frac{{\left({{\vec{P}}^{2}+{m}^{2}}\right)}^{1/2}+\left|{m}\right|-\beta
\vec{\alpha }\cdot
\vec{P}}{{\left\{{2\left[{{\left({{\vec{P}}^{2}+{m}^{2}}\right)}^{1/2}+\left|{m}\right|}\right]}\right\}}^{1/2}}}\]
and $\vec{\alpha }=\left({\begin{array}{cc}0&\vec{\alpha }\\
\vec{\alpha }&0\end{array}}\right),\qquad \beta =\left({\begin{array}{cc}I&0\\
0&-I\end{array}}\right),$ 

we obtain

\begin{equation} \left({\vec{\alpha }\cdot \vec{\beta }+\left|{m}\right|\beta }\right){\psi
}^{\left({\pm }\right)}\left({P}\right)=\pm {P}_{0}{\psi }^{\left({\pm
}\right)}\left({P}\right)\end{equation} 

The transformation properties of the Dirac vectors are

\begin{equation} U\left({a,\Lambda }\right){\psi
}^{\left({+}\right)}\left({P}\right)=\prod\limits_{\mu=0}^{3}
{\left({{\frac{1+{\frac{1}{2}}i\varepsilon {P}_{\mu }}{1-{\frac{1}{2}}i\varepsilon {P}_{\mu
}}}}\right)}^{{a}_{\mu }}\left({{\frac{1+{\frac{1}{2}}i\vec{\omega }\cdot
\vec{\sum}}{1-{\frac{1}{2}}i\vec{\omega }\cdot \vec{\sum}}}}\right) {\psi
}^{\left({+}\right)}\left({{\Lambda }^{-1}P}\right)\end{equation} for spacial rotations, and

\begin{equation} U\left({a,\Lambda }\right){\psi
}^{\left({+}\right)}\left({P}\right)=\prod\limits_{\mu=0}^{3}
{\left({{\frac{1+{\frac{1}{2}}i\varepsilon {P}_{\mu }}{1-{\frac{1}{2}}i\varepsilon {P}_{\mu
}}}}\right)}^{{a}_{\mu }}\left({{\frac{1-{\frac{1}{2}}i\vec{v}\cdot \vec{\alpha
}}{1+{\frac{1}{2}}i\vec{v}\cdot \vec{\alpha }}}}\right){\psi
}^{\left({+}\right)}\left({{\Lambda }^{-1}P}\right)\end{equation} for pure Lorentz
transformations.

The negative energy covariant vectors carry the same representation.

\vspace*{24pt}{\noindent  \bf TRANSFORMATION OF THE COVARIANT FUNCTIONS IN \\
CONFIGURATION SPACE}\vspace*{12pt} 

\setcounter{section}{6}\setcounter{equation}{0}
In order to construct the wave functions in configuration space we solve the Klein-Gordon and
Dirac wave equation on the lattice. This procedure can be easily generalized to higher spin
representations.

We define the scalar function on the $(3+1)$ dimensional cubic lattice

\[\phi \left({{j}_{1}{\epsilon }_{1},{j}_{2}{\epsilon }_{2},{j}_{3}{\epsilon }_{3},n\tau
}\right)\equiv \phi
\left({\vec{j},n}\right)\] where ${\epsilon }_{1},{\epsilon }_{2},{\epsilon }_{3},\tau$ are
small quantities in the space-time directions and
${j}_{1},{j}_{2},{j}_{3},n$ are integer numbers.

We define the difference operators

\[{\delta }_{\mu }^{+}\equiv {\frac{1}{{\epsilon }_{\mu }}}{\Delta }_{\mu }\prod\limits_{\nu
\ne \mu }^{} {\tilde{\Delta }}_{\nu },\mu ,\nu =0,1,2,3 \qquad {\delta }_{\mu }^{-}\equiv
{\frac{1}{{\epsilon }_{\mu }}}{\nabla }_{\mu }\prod\limits_{\nu \ne \mu }^{} {\tilde{\nabla
}}_{\nu }\]

\[{\eta }^{+}\equiv \prod\limits_{\mu =0}^{3} {\tilde{\Delta }}_{\mu } \qquad {\eta
}^{-}\equiv \prod\limits_{\mu =0}^{3} {\tilde{\nabla }}_{\mu }\]

Then the KLein-Gordon wave equations defined on the grid points of the lattice can be read off

\begin{equation} \left({{\delta }_{1}^{+}{\delta }_{1}^{-}+{\delta }_{2}^{+}{\delta
}_{2}^{-}+{\delta }_{3}^{+}{\delta }_{3}^{-}-{\delta }_{0}^{+}{\delta }_{0}^{-}-{M}^{2}{\eta
}^{+}{\eta }^{-}}\right)\phi \left({\vec{j},n}\right)=0\end{equation} 

It can be verified by direct substitution that the plane wave solution satisfy the diference
equation

\begin{equation} f\left({\vec{j},n}\right)\equiv {\left({{\frac{1+{\frac{1}{2}}i{\varepsilon
}_{1}{k}_{1}}{1-{\frac{1}{2}}i{\varepsilon
}_{1}{k}_{1}}}}\right)}^{{j}_{1}}{\left({{\frac{1+{\frac{1}{2}}i{\varepsilon
}_{2}{k}_{2}}{1-{\frac{1}{2}}i{\varepsilon
}_{2}{k}_{2}}}}\right)}^{{j}_{2}}{\left({{\frac{1+{\frac{1}{2}}i{\varepsilon
}_{3}{k}_{3}}{1-{\frac{1}{2}}i{\varepsilon
}_{3}{k}_{3}}}}\right)}^{{j}_{3}}{\left({{\frac{1-{\frac{1}{2}}i\tau \omega
}{1+{\frac{1}{2}}i\tau \omega }}}\right)}^{n}\end{equation}  provided the dispersion relation
is satisfied

\begin{equation} {\omega }^{2}-{k}_{1}^{2}-{k}_{2}^{2}-{k}_{3}^{2}={M}^{2}\end{equation} 

From section 5 the Klein-Gordon equation is invariant under finite Lorentz transformations.

The discrete version of the Dirac wave equation can be written as

\begin{equation} \left({{\gamma }_{1}{\delta }_{1}^{+}+{\gamma }_{2}{\delta }_{2}^{+}+{\gamma
}_{3}{\delta }_{3}^{+}-i{\gamma }_{0}{\delta }_{0}^{+}+M{\eta }^{+}}\right)\psi
\left({\vec{j},n}\right)=0\end{equation}  where ${\gamma }_{\mu }, \quad \gamma=0,1,2,3$ are
the usual Dirac matrices. Applying the operator

\[\left({{\gamma }_{1}{\delta }_{1}^{-}+{\gamma }_{2}{\delta }_{2}^{-}+{\gamma }_{3}{\delta
}_{3}^{-}-i{\gamma }_{0}{\delta }_{0}^{-}-M{\eta }^{-}}\right)\] from the left on both sides
of (6.4) we recover the Klein-Gordon equation (6.1). Let now construct solutions to (6.4) of
the form

\[\psi \left({\vec{j},n}\right)=\omega \left({\vec{k},E}\right)\ f\left({\vec{j},n}\right)\]
where the
$f\left({\vec{j},n}\right)$ are given in (6.2).

The four-component spinors $\omega \left({\vec{k},E}\right)$, with spatial momentum
$\vec{k}\equiv
\left({{k}_{1},{k}_{2},{k}_{3}}\right)$, must satisfy

\begin{equation} \left({i\vec{\gamma }\cdot \vec{k}-{\gamma }_{0}E+M}\right)\omega
\left({\vec{k},E}\right)=0\end{equation}  as in the continuous case. Multiplying this
equation from the left by

\[\left({i\vec{\gamma }\cdot \vec{k}-{\gamma }_{0}E-M}\right)\] we obtain the dispersion
relation

\begin{equation} {E}^{2}-{\vec{k}}^{2}={M}^{2}\end{equation} 

The transformation properties of the wave functions in the configuration space are given, as
in the continuous case, as follows:

\begin{equation} U\left({a,\wedge }\right)\psi \left({\epsilon
j}\right)=\left({{\frac{1+{\frac{i}{2}}i\vec{\omega }\cdot
\vec{\sum}}{1-{\frac{i}{2}}\vec{\omega }\cdot \vec{\sum}}}}\right)\psi \left({{\wedge
}^{-1}\left({\epsilon j-a}\right)}\right)\end{equation}  for spatial rotations of angle
$\vec{\omega }$ and

\begin{equation} U\left({a,\wedge }\right)\psi \left({\varepsilon
j}\right)=\left({{\frac{1-{\frac{i}{2}}\vec{u}\cdot \vec{\alpha
}}{1+{\frac{i}{2}}\vec{u}\cdot \vec{\alpha }}}}\right)\psi \left({{\wedge
}^{-1}\left({\varepsilon j-a}\right)}\right)\end{equation}  for special Lorentz
transformation of relative velocity $\overline{u}=\overline{v}/v$.

The connection of these transformations in the configuration space and the transformation in
momentum space is given via the Fourier transform in the lattice, for non-periodic functions,
namely,

\begin{equation} \psi \left({P}\right)=\sum\limits_{j=-\infty }^{\infty }
{\left({{\frac{1+{\frac{i}{2}}i\epsilon P}{1-{\frac{i}{2}}i\epsilon P}}}\right)}^{j}\psi
\left({\epsilon j}\right)\end{equation} where $\psi \left({j\epsilon}\right)$ satisfy boundary
conditions $\psi \left({\epsilon j}\right)\rightarrow 0$, when $j\rightarrow \infty$ and

\[\sum\limits_{j=-\infty }^{\infty } \psi \left({\varepsilon j}\right)<\infty \quad [12]\] 

Using sumation by parts and the boundary conditions we derive that (5.11) is the Fourier
transform of (6.7) and (5.12) is the Fourier transform of (6.8), and that (6.5) is the
Fourier transform of (6.4).

\vspace*{24pt}{\noindent  \bf ACKNOWLEDGMENT}\vspace*{12pt} 

The author wants to express his gratitude to professor Peter Kramer for many iluminating
critics and remarks.

This work has been partially supported by spanish grant of D.G.I.C.Y.T. (Project PB 94 -
1318).


\begin{thebibliography}{99}
\bibitem{} The consistency of this new discrete Fourier transform is proved with the help of
the $\delta$ sequence:

\[{s}_{L}\left({j}\right)=\sum\limits_{k=-L}^{L} {\left({{\frac{1+{\frac{i}{2}}\epsilon
k}{1-{\frac{i}{2}}\epsilon k}}}\right)}^{j}\] that satisfies $\lim\limits_{L\rightarrow
\infty }^{}\ {s}_{L}\left({0}\right)=\infty$ and


\[\lim\limits_{N\rightarrow \infty }^{}{\frac{1}{2N}}\sum\limits_{j=-N}^{N}
{s}_{L}\left({j}\right)=1\] for all L.
\end{thebibliography}
\end{document}